\documentclass[apjl]{emulateapj}

\newcommand{\noprint}[1]{}

\newcommand{\kms}{\,\mathrm{km}\,\mathrm{s}^{-1}}

\newcommand{\kyr}{\,\mathrm{kyr}}

\newcommand{\Kv}{\,\mathrm{K}}

\newcommand{\kpc}{\,\mathrm{kpc}}
\newcommand{\pc}{\,\mathrm{pc}}

\newcommand{\ergs}{\,\mathrm{erg}\,\mathrm{s}^{-1}}
\newcommand{\cmq}{\,\mathrm{cm}^{-3}}

\newcommand{\Msun}{\,\mathrm{M}_{\sun}} 
\newcommand{\Msunyr}{\,\mathrm{M}_{\sun}\,\mathrm{y}^{-1}}

\newcommand{\vrw}{\left<v_\mathrm{r,w}\right>}
\newcommand{\vrwout}{\left<v_\mathrm{r,w,out}\right>}

\newcommand{\sigmatot}{\sigma_\mathrm{tot}}
\newcommand{\sigmalos}{\sigma_\mathrm{los,45}}
\newcommand{\sigmaturb}{\sigma_\mathrm{turb}}

\newcommand{\PUFO}{P_\mathrm{UFO}}

\newcommand{\betaufo}{\beta_\mathrm{UFO}}

\newcommand{\betajet}{\beta_\mathrm{jet}}

\newcommand{\ncc}{n_\mathrm{c,0}}

\newcommand{\Ekinw}{E_\mathrm{kin,w}}
\newcommand{\Eintw}{E_\mathrm{int,w}}

\newcommand{\Mdotjet}{\dot{M}_\mathrm{jet}}
\newcommand{\Mdotufo}{\dot{M}_\mathrm{UFO}}

\newcommand{\eqref}[1]{(\ref{#1})}

\newcommand{\msigma}{$M$--$\sigma$}

\slugcomment{Accepted for publication in the Astrophysical Journal Letters on December 20, 2012.}

\shorttitle{UFO Feedback}
\shortauthors{Wagner, Umemura, \&{} Bicknell}

\begin{document}

\title{Ultra Fast Outflows: Galaxy-Scale Active Galactic Nucleus Feedback}

\author{A.~Y. Wagner\altaffilmark{1} \and M. Umemura\altaffilmark{1} \and G.~V. Bicknell\altaffilmark{2}}
\altaffiltext{1}{Center for Computational Sciences, University of Tsukuba, 1-1-1 Tennodai, Tsukuba, Ibaraki, 305-8577, Japan}
\altaffiltext{2}{Research School of Astronomy and Astrophysics, The Australian National University, ACT 2611, Australia}
\email{ayw@ccs.tsukuba.ac.jp}

\begin{abstract}
  We show, using global 3D grid-based hydrodynamical simulations, that Ultra Fast Outflows (UFOs) from Active Galactic Nuclei (AGN) result in considerable feedback of energy and momentum into the interstellar medium (ISM) of the host galaxy. The AGN wind interacts strongly with the inhomogeneous, two-phase ISM consisting of dense clouds embedded in a tenuous hot hydrostatic medium. The outflow floods through the inter-cloud channels, sweeps up the hot ISM, and ablates and disperses the dense clouds. The momentum of the UFO is primarily transferred to the dense clouds via the ram pressure in the channel flow, and the wind-blown bubble evolves in the energy-driven regime. Any dependence on UFO opening angle disappears after the first interaction with obstructing clouds. On kpc scales, therefore, feedback by UFOs operates similarly to feedback by relativistic AGN jets. Negative feedback is significantly stronger if clouds are distributed spherically, rather than in a disc. In the latter case the turbulent backflow of the wind drives mass inflow toward the central black hole. Considering the common occurrence of UFOs in AGN, they are likely to be important in the cosmological feedback cycles of galaxy formation.
\end{abstract}


\section{Introduction}
\label{s:intro}

Active Galactic Nuclei (AGN) winds have for a long time been considered an integral part of the feedback cycle of galaxy formation \citep{silk1998a,fabian1999a}. The kinetic energy fed back by the wind into the interstellar medium of the host galaxy is, in principle, sufficient to heat, disperse, and possibly unbind dense gas and therefore inhibit galaxy-wide star formation \citep{crenshaw2012a}. Such a mechanism, whereby the central supermassive black hole (SMBH) is capable of controlling the growth of its host galaxy on scales much greater than its gravitational sphere of influence is an attractive idea to explain the apparent co-evolution of SMBH and galaxy as evidenced by the \msigma{} relation \citep{ferrarese2000a,gebhardt2000a}, and the shape and evolution of galaxy luminosity functions and BH mass functions over cosmic time \citep{croton2006a,croom2009a,merloni2008b}.

Mass outflows are common in all types of AGN and, those most closely associated with disc winds are observed in UV and X-ray absorption lines \citep{crenshaw2003b}. 
An extreme class or component of outflows recently detected in highly ionized and highly blueshifted Fe K-shell absorption lines in the hard X-ray band are ultra fast outflows \citep[UFOs, ][]{cappi2006a,tombesi2010a}. UFOs are thought to be mildly relativistic disc winds, with speeds $v\sim0.01c$ to $0.1c$ (several $10^3\kms$ to $10^4\kms$), originating within 100 -- $10^4$ gravitational radii of the SMBH. From a sample of 42 local radio-quiet AGN, \citet{tombesi2010b,tombesi2012a} determined that the incidence of UFOs in AGN is greater than $40\%$, that mass outflow rates are typically $0.01$ -- $1\Msunyr$, and that the outflow kinetic power is $10^{42}$ -- $10^{45}\ergs$. These authors also suggested that such outflows ought to have a strong feedback effect on the evolution of the host galaxy.

The effectiveness of any mode of AGN feedback depends sensitively on the properties of the ISM, but there are few detailed studies incorporating realistic multi-phase distributions (\citealp{saxton2005a}; \citealp[][SB07 henceforth]{sutherland2007a}; \citealp[][WBU12 henceforth]{cooper2008a, gaibler2012a, wagner2012a}). WBU12 showed with 3D hydrodynamical simulations of relativistic AGN jets interacting with a two-phase ISM that jets with powers $10^{43}$ -- $10^{46}\ergs$ are capable of efficient energy and momentum transfer to disperse the dense gas in the bulge of galaxies to velocities commonly observed in (high-redshift) radio galaxies \citep{morganti2005a,nesvadba2008a}, and beyond those predicted by the \msigma{} relation, if the Eddington ratio of the jets is greater than $10^{-4}$. The dominant force responsible for the efficient energy and momentum transfer was identified as the ram pressure carried by jet streams that percolate the porous two-phase ISM.


Although UFOs, on average, have kinetic powers an order of magnitude less than AGN jets, their mass outflow rates are comparable to their accretion rates, and thus, for the same kinetic power, they carry considerably more momentum \citep{tombesi2012a} than jets. UFOs may therefore substantially affect the galaxy-scale ISM of the host, in particular the dense, warm and cold phases of the ISM from which stars could form. The present letter tests this proposition with global 3D hydrodynamic simulations of UFOs interacting on kpc scales with the two-phase ISM of the host galaxy.

\section{Equations, code, and initial and boundary conditions}\label{s:code}

Let $\rho$, $\mathbf{v}$, $p$, $\mathbf{I}$, $T$, $\Lambda$, $\phi$, $\gamma=5/3$, $k$, and $\mu$ be the fluid density, the three dimensional velocity vector, the pressure, the unit tensor, the temperature, the cooling rate, the gravitational potential, the adiabatic index for an ideal gas, Boltzmann's constant, and the mean mass per particle, respectively. The system of equations describing the UFO outflow, hot atmosphere, and warm clouds in the one-fluid approximation is \citep{landau1987a}:
\begin{eqnarray}
  \frac{\partial \rho}{\partial t} &+& \nabla \cdot (\rho \mathbf{v)} = 0\,;  \nonumber \\
  \frac{\partial \rho \mathbf{v}}{\partial t} &+& \nabla \cdot (\rho \mathbf{v}\mathbf{v} + p\mathbf{I}) = \rho \nabla \phi\,; \label{e:eqns}\\
  \frac{\partial}{\partial t}\left( \frac{1}{2}\rho \mathbf{v}\cdot\mathbf{v} + \frac{1}{\gamma - 1}p\right) &+& \nabla \cdot \left[\mathbf{v} \left( \frac{1}{2}\rho \mathbf{v}\cdot\mathbf{v} + \frac{\gamma}{\gamma - 1}p\right)\right] \nonumber\\
&& = -\rho^2\Lambda(T) - \rho \nabla \phi\cdot \mathbf{v}\,; \nonumber\\
&p&= \rho k T / \mu \,. \nonumber
\end{eqnarray}
We integrate these equations using the publicly available, open-source Eulerian Godunov-type code PLUTO \citep{mignone2007b} version 3.1.1.  


The UFO inlet is placed at $x=(0,0,0)$ and is directed in the positive $x$-direction with an opening angle $\theta=30^\circ$. The velocity, mass injection rate, and energy injection rate, at the inlet are $v=0.1c$, $\Mdotufo=0.1\Msunyr$, and $\PUFO=10^{44}\ergs$, respectively. These are typical UFO parameter values, and represent a case where the internal and kinetic energy injection rates are comparable. The energy partition may change during the evolution of the UFO from its origin near the accretion disc to the scale corresponding to the smallest cell resolution in our simulations ($2\pc$).

A crucial ingredient in these simulations is the two-phase ISM, which consists of a warm ($T \sim 10^4 \Kv$) phase and a hot ($T\sim 10^7 \Kv$) phase. The hot phase is in hydrostatic equilibrium in a two-component gravitation potential, described by the sum of a \citet{hernquist1990a} and a NFW \citep{navarro1996a} density profile with core densities and scale heights $(n_c, r_c)=(200\cmq,2\kpc)$ and $(3\cmq,20\kpc)$, representing the baryonic and dark matter components, respectively. 

The warm phase ISM is initialized using the algorithm by \citet{lewis2002a}, which generates a 3-dimensional random density distribution that simultaneously satisfies single-point lognormal statistics and two-point fractal statistics. These statistical properties are consistent with those observed of molecular clouds in our Galaxy \citep{kainulainen2009a,roman-duval2010a} and simulations of self-regulated multi-phase ISM turbulence \citep{wada2001a}. We choose the same statistical parameters for the fractal cube as those used by SB07.

The clouds in our simulations are initially in pressure equilibrium with the hot phase and we investigate two cases for the spatial profile of their mean density: one is that of an isothermal sphere, the other is that of a turbulent quasi-Keplerian disc (SB07). The central mean densities, $\ncc$, are $300\cmq$ and $1000\cmq$, respectively, and the total mass of clouds in both cases is $\sim4\times10^8\Msun$. Individual clouds have peak densities of a few $10^5\cmq$ and temperatures $\lesssim 100\Kv$. The disc is settled in the $y$ -- $z$ plane and centered at $x=(0,0,0)$ with scale height $h_c$ determined by the combination of the mean sound speed, $\bar{a}\approx10\kms$, and the turbulent velocity dispersion, $\sigmaturb=40\kms$, $h_c=\sqrt{(\bar{a}^2+\sigmaturb^2)/4\pi G\mu\ncc}\sim40\pc$ (SB07). Turbulent support is formally introduced to ensure a fairly large disc scale height while keeping cloud temperatures below $10^4\Kv$, so that the presence of sufficiently large, massive clouds within a scale height results in strong interactions between the wind and the ISM.

Advected scalar variables distinguish UFO material and warm phase gas from each other and from the hot phase background. We include non-equilibrium, optically thin atomic cooling for $T>10^4\Kv$ \citep{sutherland1993a}.To emulate heating due to photoionization, $\Lambda=0$ for $T<10^4$. Cloud cores initially have temperatures $\sim100\Kv$ with densities up to $\sim10^5\cmq$. Thermal conduction, photo-evaporation, self-gravity, and magnetic fields are not included.

The cubical simulation domain with physical dimensions of $1\kpc^{3}$ consists of $352\times512\times512$ cells in a cartesian grid providing a spatial resolution of approximately $2\pc$ per cell. Due to the nearly adiabatic nature of the flow, a restricted one parameter scaling of physical dimensions by up to a factor of 5 is possible (SB07).

\section{Results}\label{s:results}

\begin{figure*}
  \begin{center}
  \includegraphics[width=1.0\linewidth]{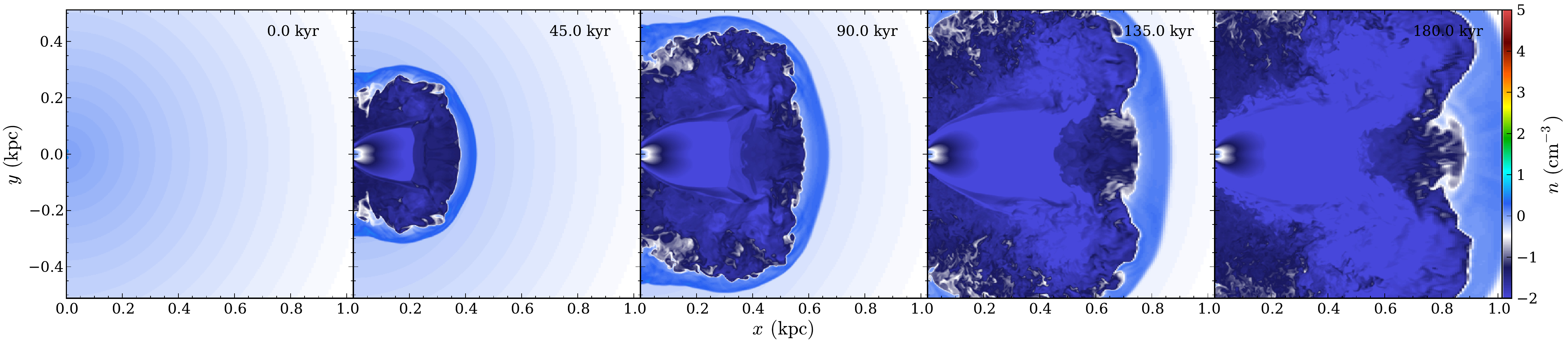}
  \caption{Midplane density slices of the evolution of a $10^{44}\ergs$ UFO in an ISM devoid of clouds (Run A). See the electronic edition of the Journal for a color version of this figure.}\label{f:rho1}
  \end{center}
\end{figure*}

\begin{figure*}
  \begin{center}
  \includegraphics[width=1.0\linewidth]{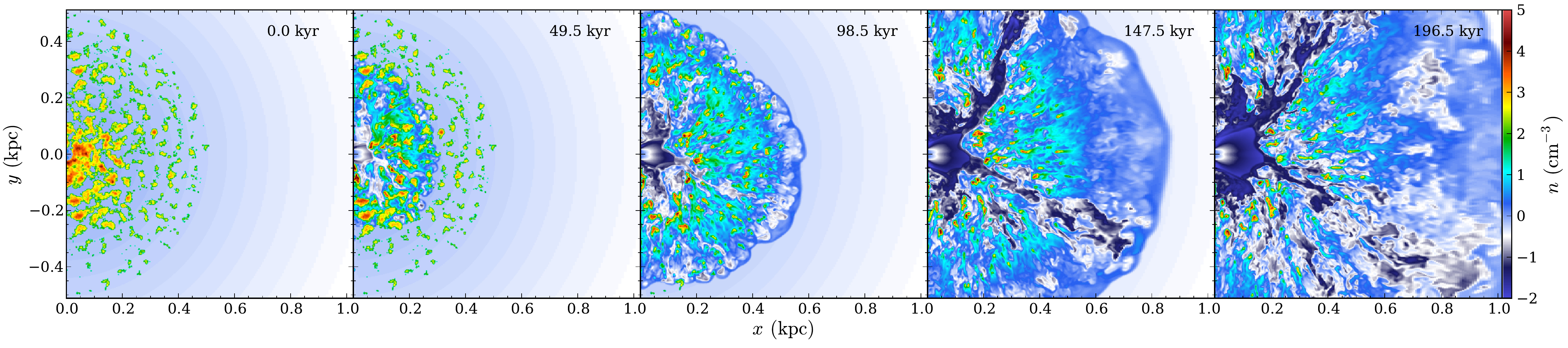}
  \caption{Same as Fig~\ref{f:rho1}, but for a two-phase ISM with spherically distributed clouds (Run B).}\label{f:rho2}
  \end{center}
\end{figure*}

\begin{figure*}
  \begin{center}
  \includegraphics[width=1.0\linewidth]{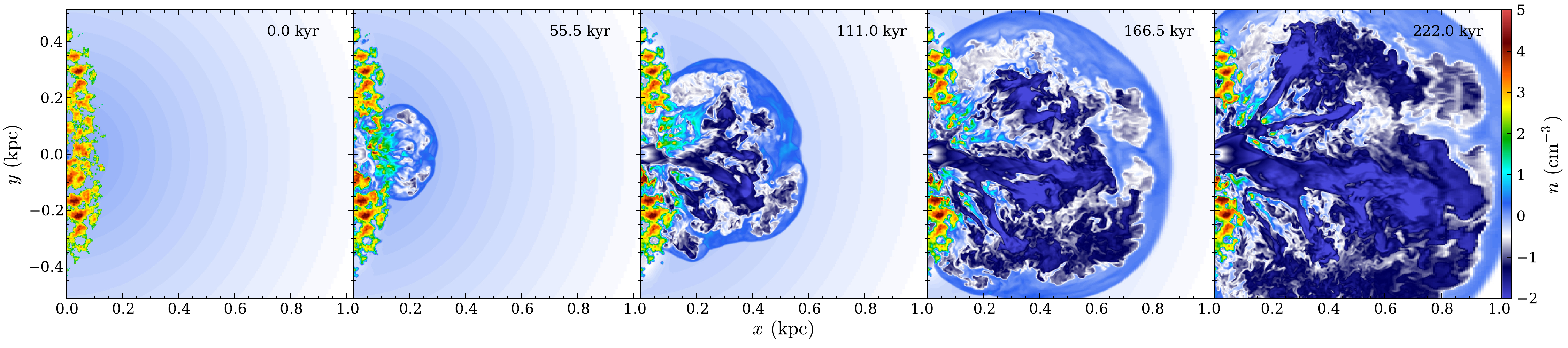}
  \caption{Same as Fig~\ref{f:rho1}, but for a two-phase ISM with clouds distributed in a quasi-Keplerian dis (Run C).}\label{f:rho3}
  \end{center}
\end{figure*}

The evolution of the UFO is shown in Figures~\ref{f:rho1}, \ref{f:rho2}, and \ref{f:rho3} for three simulations with different initial ISM. Figure~\ref{f:rho1} shows midplane density slices of the expansion of a UFO into a hot atmosphere devoid of clouds, which we shall call run A.
Figures~\ref{f:rho2}, \ref{f:rho3} show simulations of the UFO interacting with a two-phase ISM in which the clouds are distributed spherically or in a disc, referred to as run B and run C, respectively.


In run A, the UFO expands in a self-similar fashion into a single-phase, smooth, hot hydrostatic medium, giving rise to a relatively well-defined two-shock structure and unstable contact discontinuity surface \citep[c.f.][]{weaver1977a}. Confined by the shock-heated ambient medium, the turbulent flow beyond the reverse shock circulates back outside the freely expanding wind toward the plane of the galaxy.

In the common, very early stage of evolution for runs B and C, the freely expanding wind interacts strongly with the first clouds in its path and is isotropically diverted into sub-streams. Consequently, any dependence on opening angle disappears at this point. Within $10\kyr$ after the start of the UFO, however, the evolution begins to differ between the cases for bulge-like and disc-like cloud distributions. In the former, the UFO streams continue to branch out isotropically and inflate a quasi-spherical energy bubble. The flow entirely engulfs and ablates the clouds, primarily driving them radially outward in long, cometary filaments. The diffuse warm filaments reach speeds of $1000\kms$ while the colder cores are accelerated up to $200\kms$. The results from run B are similar to those reported by \citet{saxton2005a} and WBU12 for AGN jet feedback.

Because of the higher densities and filling factors along the galactic plane in run C, the secondary UFO streams do not disperse the clouds at large disc radii. As the UFO breaks out through the center of the disc carrying with it an appreciable mass of dense cloud material, it inflates an orbed energy bubble above the disc, which sweeps back down over the clouds in the outer regions of the disc. These clouds are compressed and accelerated toward the galactic plane and central BH by the turbulent, ram-pressure dominated back-flow in the bubble. The results of this run are similar to those of the simulations by SB07 and \citet{gaibler2012a} for AGN jets interacting with a dense galactic disc.

Runs B and C demonstrate that the feedback on the warm phase of the ISM depends strongly on the spatial distribution of clouds. In all runs, however, the UFO-blown bubble remains in the energy-driven regime, despite radiative cooling in the clouds. This is consistent with the predictions of recent analytic models by \citet{faucher-giguere2012a}, which also justify our neglect of inverse-Compton cooling.

\begin{figure}
  \begin{center}
  \includegraphics[width=1.0\linewidth]{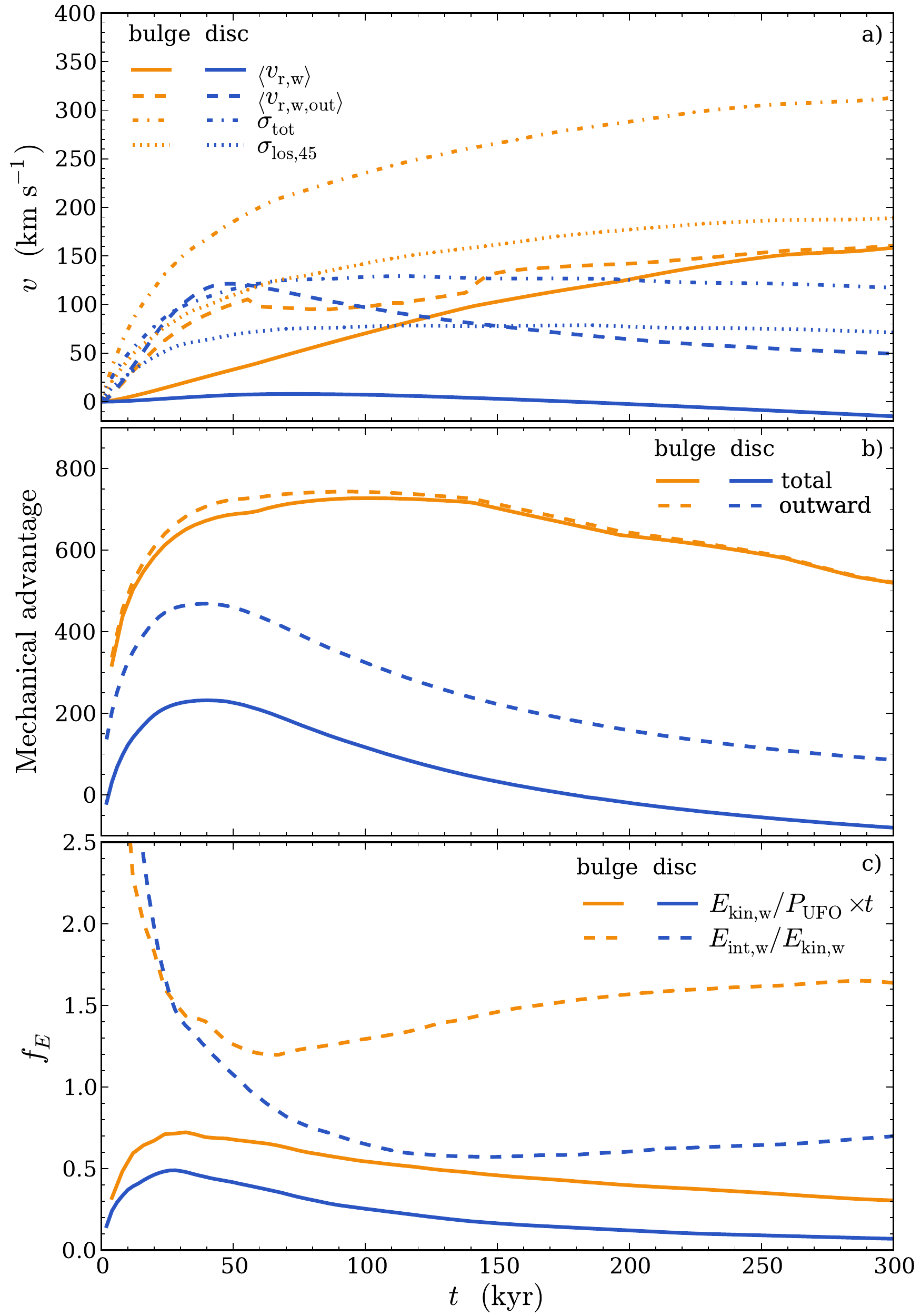}
  \caption{Evolution of various quantities that gauge the feedback efficiency for simulations with a bulge-like or a disc-like distribution of clouds: a) The density-weighted average radial velocity, the outward (positive) component of the radial velocity, the total velocity dispersion, and the line-of-sight velocity dispersion (at $45^\circ$ inclination); b) The mechanical advantage as measured by the total or outward-only radial momenta of clouds. c) The warm-phase kinetic energy as a fraction of the energy provided by the UFO and the ratio of the warm-phase internal energy to kinetic energy.}\label{f:tri}
  \end{center}
\end{figure}

In the following, we use the four quantities to measure the efficiency of feedback by the UFO: the mean radial velocity, the velocity dispersion, the mechanical advantage, and the kinetic energy of the clouds.

We define the density-weighted mean radial outflow velocity of the warm phase, $\vrw=\sum \phi_w \rho_w \mathbf{v} \cdot \mathbf{\hat{r}}/ \sum \phi_w \rho_w $ \citep{wagner2011a}. The evolution of this quantity and its outward only (positive) component are plotted together with the total velocity dispersion, $\sigmatot$, and the ($45^\circ$) line-of-sight velocity dispersion, $\sigmalos$, as a function of time in Fig.~\ref{f:tri}~a). We see that for the case of a bulge-like cloud distribution (run B), the velocities of the warm phase reach several $100\kms$, and keep increasing for the duration of the simulation. At late stages of the evolution, the clouds are predominantly accelerated outward ($\vrw\approx\vrwout$), although their radial speed never quite reaches the escape velocity of this system, which is $\sim450\kms$ at $0.5\kpc$. The velocity dispersions, however, reach values beyond those predicted by the \msigma{} relation, which for the simulated galaxy using the relations by \citet{graham2012a} and a black hole mass of $6\times10^7\Msun$ is $\sim170\kms$. The values of $\vrw$ and $\sigma$ are comparable to those found in analogous simulations of AGN jet feedback (c.f. WBU12).


In run C, the feedback in terms of radially outward directed cloud acceleration and cloud velocity dispersions is noticeably less efficient. The radial outflow velocity peaks early (after $50\kyr$) as bulk cloud material is pushed out of the galactic disc and then drops throughout the rest of the simulation as the energy of the UFO is primarily channeled into inflating the bubble beyond. Infall brought about by the surrounding overpressure and turbulent backflow dominates the dense gas motions after $200\kyr$ resulting in net accretion. The velocity dispersions also saturate well below $150\kms$ and do not reach the value predicted by the \msigma{} relation for this galaxy.

For a given kinetic power, the ratio of the mass outflow rate of the UFO to that of the jet is $\Mdotufo/\Mdotjet\sim2(\Gamma-1)/\betaufo^2\gtrsim1000$, where $\Gamma$ is the jet Lorentz factor and $\beta=v/c$. The momentum delivered by the UFO is consequently larger by a factor $\sim2(\Gamma-1)/\Gamma\betajet\betaufo\gtrsim50$. In an expanding wind, the momentum transfer leading to the acceleration of embedded clouds in all directions is provided by the sum of the ram pressure and thermal pressure integrated over the surface of the clouds. Because the surface area increases over time, this system exhibits a mechanical advantage greater than unity, and care is required when assessing momentum budgets: the net (scalar, not vector) momentum may be larger than that injected by the UFO over a given time. Panel b) in Fig.~\ref{f:tri} shows that the mechanical advantage with respect to the clouds, defined here as the ratio of the total radial momentum of the warm-phase at a given time to the total UFO momentum injected up to that time, in both runs B and C is much greater than unity, indicating efficient momentum transfer.\footnote{Note that this definition of the mechanical advantage is somewhat different to that used by other authors, e.g. \citet{faucher-giguere2012a}.}. The efficiency is higher than that for AGN jets by almost an order of magnitude (cf. WBU12).

Figure \ref{f:tri} c) shows the evolution in time of the fraction of warm-phase kinetic energy to the integrated energy injected by the UFO up to time $t$, $\Ekinw/\PUFO\times t$, and the ratio of the warm-phase internal energy to its kinetic energy, $\Eintw/\Ekinw$. In both runs B and C, the fraction of energy imparted by the jet to the warm phase peaks early at $\gtrsim50\%$ and subsequently declines slowly. Overall, however, the energy transfer efficiency, both in terms of heating and accelerating the warm phase is higher in the case of spherically distributed clouds, compared to the case of clouds distributed in a disc. Due to a higher mechanical advantage in the first $\sim100\kyr$, the energy transfer rate is somewhat higher for UFOs than for AGN jets.

\section{Discussion and Conclusions}\label{s:concl}

The simulations presented in this work confirm that powerful UFOs are capable of generating strong, galaxy-wide feedback. Energy and momentum transfer is achieved by fast, mass-entrained flows through the porous channels of the two-phase ISM, which carry high ram pressure to clouds at all locations of the galaxy, even in the plane of the disc. In the case of spherically distributed clouds, the feedback results in strongly heated and dispersed clouds, accelerated outward from the galaxy bulge. In the case of clouds distributed in a disc, the feedback results in a rapid lift-up of clouds from the plane of the disc followed by compression and net inflow of warm disc material toward the center of the galaxy. Because of the strong interaction of the freely expanding wind with the first obstructing clouds, the results do not depend on the opening angle of the UFO.

Only two ISM distributions have been presented here, whereas negative and positive feedback efficiencies depend on a range of ISM parameters. For example, for jet-mediated AGN feedback the feedback efficiency depends strongly on the mean density and mean size of clouds, but only weakly on cloud volume filling factor (WBU2012). Feedback by UFOs and AGN jets operate alike, and there are several reasons to expect this: 1) the injected powers are comparable; 2) the ram pressure carried by the fast channel flow is comparable because its density is primarily determined by that of the swept up hot phase; and 3) there is no dependence on opening angle. Given that the energy and momentum transfer mechanisms to the ISM are the same for jet- and UFO-driven feedback, it is reasonable to expect that the efficiency dependencies on ISM parameters for the two scenarios are similar. This work shows in addition that the spatial distribution of the clouds (e.g., spherical or in a disc) affects the feedback efficiency substantially.

One could now further study the dependence on the scale-height of the disc or on the original orientation of the outflow. One might expect, for example, that a higher degree of misalignment of the UFO with respect to the galactic disc normal will lead to stronger negative feedback.

The interactions between an AGN wind or jet and the ISM lead to heavily mass-loaded outflows, which is compatible with the requirement for ram-pressure dominated AGN outbursts in the simulations by \citet{gaspari2012a,gaspari2012b} of the feedback cycle that regulates the thermodynamics of cooling flow clusters. On galaxy and cluster scales, where mass-loaded jets may have decelerated to sub-relativistic velocities \citep{bicknell1984a,komissarov1994a}, the distinction between such slow, massive, wide jets \citep{sternberg2007a} and AGN winds may be of lesser importance as far as AGN feedback is concerned \citep{gaspari2012a}.

While mass outflow and infall rates can be well determined, our simulations do not contain all the necessary physics (e.g. self-gravity, molecule formation and cooling, photoionization, etc.) to deduce whether star-formation in surviving dense clouds is fully suppressed or whether in some cases, most likely within the larger cloud complexes in disc galaxies, star-formation may in fact be induced, due to the overpressurization of the ambient medium \citep[cf.][]{gaibler2012a}. The longevity and lateral contraction of filaments drawn out of the galactic plane by the wind may also become star-formation sites, a mechanism proposed recently by \citet{silk2012a} to explain Milky Way hypervelocity stars.


The boundary conditions for the inlet are an extrapolation of the observed UFO parameters, whose origins are near the central accretion disc. The evolution of the UFO from disc wind on scales $\sim10^{-4}$ -- $10^{-2}\pc$ to outflow on scales $\sim\pc$ is in need of future investigation \citep[see, e.g.,][for radiation-driven outflows]{wada2012a}. A class or component of outflows less powerful than UFOs known as warm absorbers \citep{mckernan2007a,torresi2012a} seen in soft X-ray absorption, may help constrain the modelling of the wind in this regime. It also remains to be seen with simulations of the evolution of the outflow on scales of order tens of kpc whether the filaments of dense outward-moving gas may be unbound from the galaxy potential and contribute to the enrichment of the intergalactic medium.

\acknowledgments
A subset of the computations were undertaken on the NCI National Facility at the Australian National University. The work was also supported by iVEC through the use of advanced computing resources located at iVEC@Murdoch. Ralph Sutherland provided us with code to generate the initial warm-phase distribution. This work was supported in part by the {\it FIRST} project based on Grants-in-Aid for Specially Promoted Research by MEXT (16002003) and JSPS Grant-in-Aid for Scientific Research (S) (20224002).


\bibliographystyle{apj}

\end{document}